%% file: D7.tex
\documentclass{JHEP3}\newcommand{\format} {\JHEPformat}

\usepackage{amsmath}
\usepackage{epsfig}
\usepackage{latexsym}

\newcommand{\JHEPformat} {
\bibliographystyle{JHEP}
\newcommand{\maketitlepage} {}
\abstract{\theabstract}
\keywords{\thekeywords}
\preprint{\thepreprint}
}

\newcommand{\TITLE}[1] {\newcommand{\thetitle} {#1}\title{#1}}
\newcommand{\ABSTRACT}[1] {\newcommand{\theabstract} {#1}}

\newcommand{\ADDRESS}[1] {\newcommand{\theaddress} {#1}}
\newcommand{\DATE}[1] {\newcommand{\thedate} {#1}\date{#1}}
\newcommand{\KEYWORDS}[1] {\newcommand{\thekeywords} {#1}}
\newcommand{\PREPRINT}[1] {\newcommand{\thepreprint} {#1}}

\newcommand{\be}{\begin{equation}}
\newcommand{\ee}{\end{equation}}

\newcommand{\p}{{\partial}}
\newcommand{\half}{{\frac{1}{2}}}

\newcommand{\diff}{{\textrm{d}}}

\TITLE{Meson spectroscopy from holomorphic probes on the warped deformed conifold}
\ADDRESS{School of Physics and Astronomy\\
  The Raymond and Beverly Sackler Faculty of Exact Sciences\\
  Tel Aviv University, Ramat Aviv, 69978, Israel
}

\author{
Stanislav Kuperstein\\
School of Physics and Astronomy\\
The Raymond and Beverly Sackler Faculty of Exact Sciences\\
Tel Aviv University, Ramat Aviv, 69978, Israel.\\
E-mail:
\email{kupers@post.tau.ac.il}
}

\ABSTRACT{

We study D$7$ brane probes holomorphically embedded in the Klebanov-Strassler model.
Analyzing the $\kappa$-symmetry condition for
D$7$ branes wrapping a $4$-cycle of a deformed conifold we find 
configurations that do not break $\mathcal{N}=1$ supersymmetry of the background.
We compute the fluctuations of the probe around one of these configurations
and obtain the spectrum of vector and scalar flavored mesons in the dual gauge theory.
The spectrum is discrete and exhibits a mass gap. 
}

\DATE{November 2004}
\KEYWORDS{The AdS/CFT correspondence}
\PREPRINT{TAUP-2789-04\\{\tt hep-th/0411097}}

\format

\begin{document}

\maketitlepage

\section{Introduction}

The AdS/CFT correspondence \cite{Maldacena:1998re},
\cite{Gubser:1998bc}, \cite{Witten:1998qj} (see
 \cite{Aharony:1999ti} for a review)
is a conjectured equivalence between a
string theory (type IIB on $AdS_5 \times S^5$) and a gauge theory
($\mathcal{N}=4$ $SU(N_{\textrm{c}})$ SYM in four dimensions). 
In the the large 't Hooft coupling limit ($g^2_{\textrm{YM}} N \to \infty$)
we can neglect the string massive modes
using type IIB supergravity on the string side of the correspondence.

Since the formulation of
the AdS/CFT correspondence its extension to 
theories with less
supersymmetries and with no conformal invariance has been of great interest.
In particular, there are well known examples of supergravity duals of $\mathcal{N}=1$  
gauge theories \cite{Maldacena:2000yy}, \cite{Klebanov:2000hb}
(see \cite{Bigazzi:2003ui}, \cite{Bertolini:2003iv}  and \cite{Aharony:2002up} for a review).
Unfortunately, most of the known supergravity backgrounds do not include
quarks in the fundamental representation. 
In the large $N$ limit the Feynman diagrams of Yang-Mills theory 
are related to a genus expansion of closed strings.
Adding fundamental flavors in the gauge theory introduces boundaries
in this expansion. It means that on the string side we have to incorporate D$p$ brane.
If the number of these branes  $N_{\textrm{f}}$ is much smaller
than $N_{\textrm{c}}$ we can neglect the back-reaction of the brane probe on the 
supergravity bulk geometry.

Adding fundamental flavors in the gauge theory requires adding of $4d$ space-time
filling D$p$ branes in the bulk.
In order to avoid a problem with tadpoles the net charge of a space-time filling brane
has to cancel. On the other hand, the RR charge is necessary for the brane stability. 
It was proposed in \cite{Karch:2002sh} (see also \cite{Karch:2002xe}) 
that  one can solve this puzzle by having a supersymmetric
D$p$ brane wrapping a topologically trivial cycle in the internal space.
The configuration is stable provided that the masses of the modes controlling
the slipping of the probe 
off the cycle are above the BF bound.
This idea was further explored by \cite{Kruczenski:2003be} for a D$7$ probe on $AdS_5 \times S^5$.
The spectrum of mesons with arbitrary $R$-charge in $\mathcal{N}=2$ SYM 
was extracted from the masses of the probe fluctuations. 
It was shown that the spectrum is discrete with a mass gap.
For related works, see \cite{Bertolini:2001qa}, \cite{Marotta:2002gc}, 
\cite{Babington:2003vm}, \cite{Wang:2003yc}, \cite{Kruczenski:2003uq}, \cite{Evans:2004ia}, 
\cite{Barbon:2004dq}, \cite{Burrington:2004id}, \cite{Erdmenger:2004dk}.

The main purpose of this paper is to explore supersymmetric space-time filling D$7$ brane
probes in the Klebanov-Strassler (KS) model \cite{Klebanov:2000hb}.
In this supergravity background 
the metric has a standard D$3$-form and 
the internal part of the metric is given by the $6d$ deformed conifold defined by:

\begin{equation}   \label{eq:DefConifold}
\sum_{i=1}^{4} z_i^2 = \epsilon^2.
\end{equation} 
The background  involves also fractional D$5$-branes wrapped around
a shrinking $S^2$ of the conifold (see \cite{Loewy:2001pq}
for the correspondence between various gauge
theory objects and wrapped D$p$ brane probes in this model). 
It was argued in \cite{Karch:2002sh} that making the quark mass very large should 
decouple the quarks from the IR theory. Therefore the D$7$ brane probe should 
be space-time filling in UV, while ending at a finite distance at the radial
direction to be absent in the IR. This distance corresponds to
the mass of the quarks in the gauge theory and 
should appear as a free parameter in a solution of the probe equations of motion.
It was suggested in \cite{Karch:2002sh} that in the KS model the following configuration
yields a supersymmetric D$7$ embedding:

\begin{equation}    \label{eq:Karch}
z_3^2 + z_4^2= \lambda^2.
\end{equation}
Here $\lambda^2$ is a constant parameter. 
Under T-duality this is mapped into type IIA D$6$ branes \cite{Gubser:1998fp}.
In this paper we will demonstrate that this embedding indeed satisfies the $\kappa$-symmetry 
condition and therefore solves the equations of motion of the space-filling D$7$ probe.

A different D$7$ probe configuration was studied in \cite{Sakai:2003wu}. 
In terms of the coordinates $z_i$'s it may be written as:

\begin{eqnarray}
\textrm{Im} \left( z_1^2 + z_2^2 \right) = 0.
\end{eqnarray}
It was explicitly shown that this embedding solves
the probe equations of motion.
Furthermore, computing the quadratic fluctuations of the probe the authors of \cite{Sakai:2003wu}
found the masses of vector and pseudo-scalar mesons in the dual confining gauge theory. 
Although the embedding considered in this paper is not holomorphic and 
therefore breaks supersymmetry, the solution is stable and the meson spectra exhibit
a mass gap. The solution has no free mass parameter and the brane 
extends to the IR region. 

For $\epsilon=0$ the equation (\ref{eq:DefConifold}) defines a cone over $T_{1,1}$
(the conifold) and the KS solution reduces to the so called Klebanov-Tseytlin (KT) solution 
\cite{Klebanov:2000nc},
which has a singularity at the tip of the conifold.
The holomorphically embedded D$7$ in this background was considered in \cite{Ouyang:2003df}. 
In terms of the coordinates used in (\ref{eq:DefConifold}) the probe discussed
in this paper is given by:

\begin{equation}  \label{eq:Ouyang}
z_3+ i z_4=m^2,
\end{equation} 
where $m$ was identified in  \cite{Ouyang:2003df} as the quark mass. The supersymmetric supergravity
solution for the leading back-reaction effects 
was also found. This approximation, however, is valid only in the UV region.

Without fractional branes the KT solution reduces further to the Klebanov-Witten solution.
In this model the stack of D$3$ branes warps the conifold metric and the resulting
geometry is just $AdS_5 \times T_{1,1}$. Following the ideas of \cite{Witten:1998xy},
\cite{Polchinski:2000uf}
it was argued in \cite{Gubser:1998fp} that 
D$3$ branes wrapping $3$-cycles of $T_{1,1}$ are dual to dibaryon operators in the gauge theory
(see also \cite{Gukov:1998kn},\cite{Beasley:2002xv},\cite{Berenstein:2002ke}). Furthermore, a D$5$ 
brane wrapping $2$-cycles of $T_{1,1}$
behaves as a domain wall in $AdS_5$. The idea to use D$7$ branes to add fundamental quarks
to the background was proposed by \cite{Karch:2002sh} 
(see \cite{Arean:2004mm} for the recent progress).

Recently the authors of \cite{Nunez:2003cf} found a rich family of supersymmetrically
embedded D$5$-branes
in the Maldacena-Nunez (MN) background \cite{Maldacena:2000yy}. 
This was achieved by solving equations arising from the $\kappa$-symmetry condition.
The solutions involve a free parameter related to the mass of the fundamental quarks.
Moreover, exploring the quadratic fluctuations around these brane configurations admits a 
quite simple relation between the meson masses and the mass of the fundamentals.

In this paper will study space-filling supersymmetric D$7$ probes embedded in the KS solution. 
Similarly to \cite{Nunez:2003cf} our main technique is $\kappa$-symmetry.
Throughout this paper we will mostly study the embedding defined by:

\begin{equation}  \label{eq:us}
z_4 = \mu, 
\end{equation}
but we will also argue that the probe configurations  (\ref{eq:Karch})
and  (\ref{eq:Ouyang}) do not spoil supersymmetry. More generally we will show that the
$\kappa$-symmetry condition is fulfilled provided that the embedding can be written in the form
$z_4 = f(z_1^2 + z_2^2)$, where $\varphi$ is an arbitrary holomorphic function.
 This trivially holds for (\ref{eq:Karch}),
(\ref{eq:Ouyang}) and (\ref{eq:us}).
Remarkably, the embedding (\ref{eq:us}) satisfies the condition of
\cite{Karch:2002sh}. Indeed, we will demonstrate that for $\mu > \epsilon$
the D$7$ probe ends at a finite distance from the IR. 
We preferred to explore the embedding (\ref{eq:us}), since this is the only configuration
that leaves an $SO(3) \subset SO(4)$ unbroken, so that the coordinate choice
becomes relatively simple.
Using this coordinates we were able to find the spectra of the vector and scalar
mesons for the special case $\mu=0$. The spectra exhibit a mass gap and the mass
scale is that of the glueballs.

The organization of the paper is as follows. 
In Section \ref{section:KS} we give a short review of the deformed conifold and the KS model.
We find it useful to re-write the metric and the forms of the model in
terms of the $SU(2)$ invariant $1$-forms used in \cite{Papadopoulos:2000gj}. 
Based on these $1$-forms we find a convenient coordinate parameterization, 
which properly describes the embedding (\ref{eq:us}).
We also comment on the geometrical structure of the probe configuration 
for different $\mu$'s.
In Section \ref{section:Kappa} we use the $SO(4)$ invariant representation 
of the $B$-field and the metric to argue that the configurations
(\ref{eq:Karch}), (\ref{eq:Ouyang}) and (\ref{eq:us}) satisfy the
$\kappa$-symmetry condition originally derived in \cite{Marino:1999af}.
In Section \ref{section:DBIWZ} we present the D$7$ probe action
in the case of the embedding (\ref{eq:us}) with $\mu = 0$.
As usual the action includes the DBI and WZ parts. 
We then compute the vector and the scalar meson spectra  by solving
the equations for the brane quadratic fluctuation.  
The set of fields appearing in the action include the scalar
field controlling the brane fluctuation around the surface $z_4=0$,
gauge fields directed along the $4$-cycle wrapped by the brane and 
gauge fields directed parallel to the $4d$ boundary of the $10d$ metric.
We will see that there is a mixing between the scalar and one of the gauge fields in
the equations of motion.
In Section \ref{section:discussion} we summarize our results and discuss various open questions.
We summarize some useful formulae in the Appendix.

\section{The geometry of a deformed conifold and the Klebanov-Strassler model}
\label{section:KS}

In this section we will describe the deformed conifold geometry and
the KS background using an alternative formulation,
which is advantageous for the description of the embedding $z_4 = \mu$.
We will also comment on the related deformation in the dual gauge theory.

\subsection{The deformed conifold and the embedding $z_4=\mu$}

The deformed conifold is defined by (\ref{eq:DefConifold}).
It can conveniently be re-written as:

\begin{equation}
\textrm{det} W = -\epsilon^2
\qquad
\textrm{where}
\qquad
W = 
\left(
\begin{array}{cc}
z_3 + i z_4   &  z_1 - i z_2   \\
z_1 + i z_2 & -z_3 + i z_4
\end{array}
\right).
\end{equation}
In term of this matrix the radial coordinate $r$ along the conifold is given by:

\begin{equation}
r^2 = \half \textrm{Tr} \left( W^\dagger W \right) = \sum_{i=1}^4 \vert z_i \vert^2.
\end{equation}
In this paper we will use the standard re-definition of the radial coordinate
according to $r^2 = \epsilon^2 \cosh \tau$. The minimal value of the radial
coordinate is $r_{\textrm{min}} = \epsilon$ (or $\tau=0$). It corresponds to the "tip",
where the $S^2$ shrinks to zero. 
If we identify $\mathcal{F}=\mathcal{F}(r^2)$ with the K\"ahler potential, then the
metric on the deformed conifold is \cite{Candelas:1989js}:

\begin{equation}
\diff s_6^2 = \p_\alpha \bar{\p}_\beta \mathcal{F} \diff z_\alpha \diff \bar{z}_\beta
= \frac{1}{4} \mathcal{F}^{\prime\prime} \left\vert \textrm{Tr} \left( W^\dagger \diff W \right) \right\vert^2
  +  \frac{1}{2} \mathcal{F}^{\prime} \textrm{Tr} \left( \diff W^\dagger \diff W \right) .
\end{equation}
The following parameterization is used
to express the metric in terms of the radial and five angular coordinates:

\begin{equation}  \label{eq:WW0}
W = L_1 W^{(0)} L_2^\dagger 
\qquad
\textrm{where}
\qquad
W^{(0)} \equiv  
\left(
\begin{array}{cc}
0   &  \epsilon e^{\tau/2}  \\
\epsilon e^{-\tau/2}  & 0
\end{array}
\right)
\end{equation}
and the $SU(2)$ matrices $L_1$ and $L_2$ depend only on the angular coordinates:

\begin{equation}
L_i = 
\left(
\begin{array}{cc}
\cos \frac{\theta_i}{2} e^{\frac{i}{2}(\psi_i + \phi_i)}   & 
          -\sin \frac{\theta_i}{2} e^{-\frac{i}{2}(\psi_i - \phi_i)}  \\
\sin \frac{\theta_i}{2} e^{\frac{i}{2}(\psi_i - \phi_i)}   & 
          \cos\frac{\theta_i}{2} e^{-\frac{i}{2}(\psi_i + \phi_i)}  \\
\end{array}
\right)
\qquad
\textrm{where}
\qquad 
i=1,2.
\end{equation}
Since the matrix $W$ depends only on the sum $\psi_1 + \psi_2$ one usually sets
$\psi_1= \psi_2= \half \psi$.
Next we introduce the $SU(2)$ left invariant $1$-forms $\left\{ h_i, \tilde{h}_i\right\}_{i=1,2,3}$
related to the matrices $L_{1,2}$ through:

\begin{equation}    \label{eq:LdL}
L_1^\dagger \diff L_1 = \frac{i}{2} h_i \sigma^i
\qquad
\textrm{and}
\qquad
L_2^\dagger \diff L_2 = \frac{i}{2} \tilde{h}_i \sigma^i,
\end{equation}
where $\sigma_{i=1,2,3}$ are the Pauli matrices. 
The  explicit formulae for the forms in terms of the angular coordinates appear
in Appendix \ref{01}. In Appendix \ref{02} we show how these forms are connected to 
the forms $g_i$'s used in \cite{Klebanov:2000hb}.
These $1$-forms satisfy the 
$SU(2) \times SU(2)$ Maurer-Cartan
equations:

\begin{equation}    \label{eq:Maurer-Cartan}
\diff h_i = \half \epsilon_{ijk} h_j \wedge h_k,
\qquad \qquad
\diff \tilde{h}_i = \half \epsilon_{ijk} \tilde{h}_j \wedge \tilde{h}_k.
\end{equation}
Requiring the metric to be Ricci flat leads to a differential equation 
for the K\"ahler potential $\mathcal{F}(r^2)$ as described in \cite{Candelas:1989js} 
(see also(\cite{Minasian:1999tt}, \cite{Ohta:1999we}). 
In terms of the $1$-forms the 
metric of the deformed conifold takes the following form:

\begin{eqnarray} \label{eq:DeformedConifoldMetric}
\epsilon^{-\frac{4}{3}} ds_6^2 &=&
        B^2(\tau) \left( \diff \tau^2 + \left( h_3 + \tilde{h}_3 \right)^2 \right)+
    \nonumber \\ 
&& \qquad 
      + A^2(\tau) \left(  h_1^2 + h_2^2 + \tilde{h}_1^2 + \tilde{h}_2^2  +
      \frac{2}{\cosh \tau} \left( h_2 \tilde{h}_2 - h_1 \tilde{h}_1 \right) \right) ,
\end{eqnarray}
where

\begin{equation}     \label{eq:AB}
A^2 (\tau) = \frac{2^{-\frac{1}{3}}  }{4 } 
       \coth \tau \left( \sinh (2 \tau) - 2 \tau \right)^{\frac{1}{3}}
\quad
\textrm{and}
\quad
B^2 (\tau) =  \frac{2^{\frac{2}{3}} }{6 }  
    \frac{\sinh^2 \tau}{\left( \sinh (2 \tau) - 2 \tau \right)^{\frac{2}{3}}}.
\end{equation}
Let us now demonstrate how using the $1$-forms $\left\{ h_i, \tilde{h}_i\right\}$
one can parameterize the surface $z_4 = \mu$ embedded in the deformed conifold.
Let us start from the $\mu=0$ case. Since $z_4 = \frac{1}{2 i} \text{Tr} W$ and 
$\textrm{Tr} W^{(0)} = 0$ the job is done by an identification of 
the $SU(2)$ matrices $L_1$ and $L_2$:

\begin{equation}
L_1 = L_2.
\end{equation}
In terms of the angular coordinates it implies that
$\theta_1 = \theta_2$, $\phi_1 = \phi_2$,
so that  $h_i = \tilde{h}_i$ for $i=1,2,3$.  
We obtain a $4d$ surface with $z_4 = 0$ along it.
For $\mu \neq 0$ we set 

\begin{equation}   \label{eq:LLS}
L_2 = L_1 S 
\qquad
\textrm{with}
\qquad
S=
\left(
\begin{array}{cc}
\cos \frac{\gamma}{2} e^{i \frac{\delta}{2}}  & 
         - i  \sin \frac{\gamma}{2} e^{-i \frac{\delta}{2}} \\
 - i  \sin \frac{\gamma}{2} e^{i \frac{\delta}{2}} & 
          \cos \frac{\gamma}{2} e^{-i \frac{\delta}{2}}  \\
\end{array}
\right).
\end{equation}
Substituting this into (\ref{eq:WW0}) we obtain:

\begin{equation}
z_4 = \frac{1}{2 i}\textrm{Tr} W 
   = \epsilon \sin \frac{\gamma}{2} \cosh \left(\frac{\tau}{2} + i \frac{\delta}{2} \right).
\end{equation}
It means that for real $\mu$ the embedding $z_4 = \mu$ corresponds to:

\begin{equation}    \label{eq:GammaTau}
\sin \left( \frac{\gamma (\tau)}{2} \right) = \frac{\mu}{\epsilon} \frac{1}{\cosh \frac{\tau}{2}}
\qquad 
\textrm{and}
\qquad 
\delta = 0.
\end{equation}
We see that for $\mu > \epsilon$ the minimal value of the radial coordinate $\tau$
along the surface $z_4 = \mu$ is 
$\tau_{\textrm{min}} = 2 \cosh^{-1} \left(\frac{\mu}{\epsilon} \right)$
and the D$7$ brane probe embedded in this way does not reach the "tip" located at $\tau=0$.
On the other hand, for $\mu < \epsilon$ we have $\tau_{\textrm{min}}=0$.
This is shown schematically on Fig. \ref{fig}.
One can arrive at the same result directly from the definition of the deformed conifold
(\ref{eq:DefConifold}). Indeed, along the surface $z_4=\mu$ the radial coordinate
$r$ satisfies:

\FIGURE[t]{
 \label{fig}
\centerline{\input{D7KS1.pstex_t} \qquad \qquad \qquad \qquad \input{D7KS2.pstex_t}}
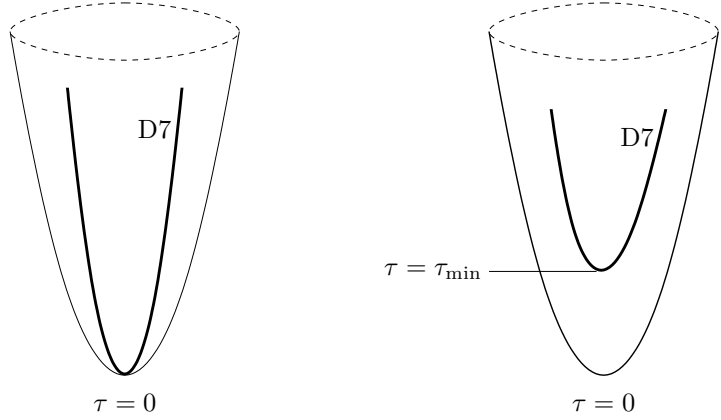
\caption{The D$7$ probe configuration $z_4=\mu$ on a deformed conifold for
$\mu < \epsilon$ (left) and for $\mu > \epsilon$ (right).
} 
}

\begin{equation}   \label{eq:rmin}
r^2 = \sum_{i=1}^3 \vert z_i \vert^2 + \mu^2 \ge  \left\vert \sum_{i=1}^3 z_i \right\vert^2 + \mu^2 
    = \left\vert \epsilon^2 - \mu^2 \right\vert + \mu^2 =
    \left\{
      \begin{array}{ll} \epsilon^2 & \quad \textrm{for} \quad \mu \le \epsilon \\
                        2 \mu^2 - \epsilon^2 & \quad \textrm{for} \quad \mu \ge \epsilon 
      \end{array}
    \right. .
\end{equation} 
Recalling that $r^2 = \epsilon \cosh \tau$ we see this is identical to (\ref{eq:GammaTau}).
It turns out that the same conclusion holds for the configurations (\ref{eq:Karch})
and (\ref{eq:Ouyang}). For instance, with
$\mu$ replaced by $\lambda$ the relation (\ref{eq:rmin}) 
holds for the case (\ref{eq:Karch}).
Notice that this result matches the prediction of (\ref{eq:Karch}).
Indeed, we see that the D$7$ brane configuration has a free parameter $\mu$
and making this parameter large enough ($\mu > \epsilon$) we find that the probe brane ends 
at a finite distance at the radial coordinate $\tau$ ($\tau_{\textrm{min}} > 0$)
and so is absent in the IR.

Using the relation (\ref{eq:LLS}) we can express the forms $\tilde{h}_1$, $\tilde{h}_2$ and
$\tilde{h}_3$ in terms of the forms $h_1$, $h_2$ and
$h_3$ and the coordinates $\gamma$ and $\delta$. Substituting (\ref{eq:LLS})
into the second equation in (\ref{eq:LdL}) we get:

\begin{eqnarray}      \label{eq:hihi}
\tilde{h}_1 &=& \left( h_1 - \diff \gamma \right) \cos \delta -
        \left( h_3 \sin \gamma + h_2 \cos \gamma \right)\sin \delta   \nonumber\\
\tilde{h}_2 &=& \left( h_1 - \diff \gamma \right) \sin \delta +
        \left( h_3 \sin \gamma + h_2 \cos \gamma \right)  \cos \delta  \nonumber\\
\tilde{h}_3 &=& \left( h_3 \cos \gamma - h_2 \sin \gamma \right) + \diff \delta         
\end{eqnarray}
As a consistency check one may verify that the forms $\tilde{h}_i$'s
satisfy the Maurer-Cartan equations (\ref{eq:Maurer-Cartan}).

The D$7$ brane in our setup spans the world-volume of the D$3$
branes and wraps the $4$-cycle defined by $z_4=\mu$. For this configuration
the coordinates $\gamma$ and $\delta$ satisfy (\ref{eq:GammaTau}).
We therefore will refer to the coordinates $x_\mu$, $\tau$ and $h_{i=1,2,3}$ 
as the world-volume coordinates
of the branes, while the coordinates $\gamma$ and $\delta$ 
will be regarded as the transverse coordinates.
Note again that the coordinate $\gamma$ is constant along the brane only for $\mu=0$.

Let us end this section with a remark on the geometry 
of the embedding $z_4 = \mu$. For $\mu \neq 0$ the configuration admits 
the $SU(2)$ isometry, which is the subgroup of the isometry group 
$SO(4) =SU(2) \times SU(2)$ of the deformed conifold.
This is similar to the isometry subgroup $SU(2) \subset SO(4)$ preserved by the 
D$7$ probe considered in \cite{Sakai:2003wu}. 
It is therefore attempting
to compare these two configurations. For $\epsilon \to 0$ the type 
IIA dual picture of the singular locus of the conifold takes the form
of a pair of perpendicular NS$5$ branes \cite{Ooguri:1995wj}, \cite{Uranga:1998vf}
\cite{Dasgupta:1998su}. 
For the deformed conifold ($\epsilon  \neq 0$) there is a "diamond"
structure at the intersection of the NS$5$ branes \cite{Aganagic:1999fe}.
Under T-duality the fractional
D$3$ branes of the KS background are transformed to D$4$ branes connecting the 
NS$5$ branes. Adding D$6$ branes on the top of the NS$5$ branes one 
introduces into the setup flavored fundamental quarks, which correspond to the strings
between the D$6$ and the D$4$ branes \cite{Elitzur:1997fh}. In the T-dual picture it gives D$7$
branes that intersect the singular locus.
The transverse coordinates of the D$7$ brane are the $S^2$ coordinates of the base.
Using the deformed conifold formulation of \cite{Gimon:2002nr} 
this structure was explicitly realized
in \cite{Sakai:2003wu}. It was further shown that the configuration
has a form of two D$7$ branes given by 
two cylinders smoothly intersecting each other at a circle at $\tau=0$.
This circle is embedded in the $S^3$ and corresponds to the "diamond".
Apart from the $SU(2)$ isometry 
preserved by the configuration there is a residual 
$U(1)$ symmetry one gets after placing
the pair of the D$7$ branes at the poles of the $S^2$.
This $U(1)$ is not a symmetry of the full background,
but rather of the induced metric on the D$7$ brane probe.
For the embedding $z_4 = \mu$ it is parallel to the $U(1)$ isometry of the induced metric
we obtain in the special case $\mu=0$. In terms of the angular coordinates introduced above
this isometry corresponds to the invariance under $\delta \to \delta + \textrm{const}$.
In our setup, however, we cannot associate directly the 
transverse coordinates ($\gamma$ and $\delta$) introduced in this section
to the coordinates of the $S^2$ and the $1$-forms $h_i$'s to the the forms of the $S^3$.
To see this, let us consider pullback of the metric (\ref{eq:DeformedConifoldMetric})
in the case of the embedding $z_4=\mu$ for $\mu=0$.  
According to (\ref{eq:hihi}) this embedding corresponds to $h_{i}=\tilde{h}_{i}$ 
and near $\tau=0$ we obtain:

\begin{equation} 
ds_6^2 = \frac{\epsilon^{4/3}}{2^{5/3}3^{1/3}} \left(
    \diff \tau^2 + 4 (h_3^2  +  h_2^2 ) \right) + O(\tau^2).
\end{equation}
We see that it does not reproduce the metric of the non-shrinking $S^3$.
We therefore cannot use directly
the "diamond" structure to argue that our D$7$ brane probe 
gets mapped into the type IIA D$6$ brane.
On the other hand, as we will discuss in the next section
the holomorphic structure of the embedding is necessary to 
preserve the supersymmetry of the KS solution.
The only holomorphic embedding with the $SU(2)$
isometry is $z_i = \textrm{const}$.
It will be very interesting to understand 
how this holomorphic embedding posses 
the T-dual description using the "diamond" structure
of the conifold similarly to \cite{Sakai:2003wu}.

\subsection{The $10d$ metric and the forms in the KS background}

Now we are in a position to present the constituents  of
the KS background in terms of the radial coordinate $\tau$,
the forms $h_{i=1,2,3}$ and the coordinates $\gamma$ and $\delta$.
The $10d$ metric
and the $5$-form flux in the KS solution have the structure of the D$3$ brane solution, namely:

\begin{equation}              \label{eq:metric}
\diff s^2 = h^{-1/2} \left( \diff x_0^2 + \ldots + \diff x_3^2 \right) +
       h^{1/2} \diff s^2_{M_6}
\end{equation}
and
\begin{equation}            \label{eq:5form}
\tilde{F}_5 = \frac{1}{g_s} (1 + \star_{10}) \diff h^{-1} \wedge \diff x_0 \wedge \ldots
                               \wedge \diff x_3,
\end{equation}
where $M_6$ is the deformed conifold metric (\ref{eq:DeformedConifoldMetric}) and the harmonic
function $h$ depends only on the coordinate $\tau$:

\begin{equation}     \label{eq:I}
h(\tau) = \frac{\epsilon^{-4/3}}{m_{\textrm{gb}}^2} I(\tau),
\qquad
\textrm{where}
\quad
I(\tau) = \int_\tau^\infty dx \frac{x \coth x -1}{\sinh^2 x} (\sinh(2x)-2x)^{1/3}.
\end{equation}
Here $m$ is the scale mass of glueballs (see \cite{Krasnitz:2000ir} 
and \cite{Caceres:2000qe}) given by

\begin{equation}      \label{eq:gb}
m^2_{\textrm{gb}} = \frac{\epsilon^{4/3}}{2^{2/3} (g_s M \alpha^\prime)^2},
\end{equation}
where $M$ is the number of the fractional D$3$ branes wrapping the shrinking $S^2$.
The five form $\tilde{F}_5$ in (\ref{eq:5form}) related to the RR $4$-form $C_4$ by

\begin{equation}
\tilde{F}_5 = F_5 + B_2 \wedge \diff C_2,
\qquad
\textrm{where}
\qquad
F_5 = \diff C_4.
\end{equation}
Here $C_2$ is the RR $2$-form.
In our case the RR $4$-form is defined by:

\begin{equation}   \label{eq:C4}
C_4 = \frac{1}{g_s} h^{-1} \diff x_0 \wedge \ldots\wedge \diff x_3
\end{equation}
Next we consider the NS $B$-field. In terms of the $1$-forms $h_i$'s and the coordinates
$\gamma$ and $\delta$ it reads:

\begin{eqnarray}  \label{eq:Bfield}
B_2 &=& \frac{1}{4} M \alpha^\prime (f-k) \bigg[
  \cosh \tau \Big(
           \left( 1 -\cos (\gamma) \right) h_1 \wedge h_2 - \sin (\gamma) h_3 \wedge \diff \gamma -
  \nonumber \\         
  &&                    - \sin (\gamma) h_1 \wedge h_3 - \cos (\gamma) h_2 \wedge \diff \gamma 
            \Big)      
    + \Big( 
          \left( \sin (\delta) h_1 + \cos (\delta) h_2 \right) \wedge \left( h_1 -\diff \gamma \right) +
  \nonumber \\ 
   &&           + \left( \cos (\delta) h_1 - \sin (\delta) h_2 \right) 
                           \wedge \left( \sin (\gamma) h_3 + \cos (\gamma) h_2 \right)
   \Big)
                                               \bigg] ,                    
\end{eqnarray} 
where 

\begin{equation}
f(\tau) = \frac{\tau \coth \tau -1}{2 \sinh \tau}(\cosh \tau - 1) ,
\qquad
k(\tau) = \frac{\tau \coth \tau -1}{2 \sinh \tau}(\cosh \tau + 1) .
\end{equation}
We learn from (\ref{eq:Bfield}) that for $\mu = 0$ (or equivalently for $\gamma,\delta=0$) 
the $B$-field vanishes along the 
brane probe.
Similarly to (\ref{eq:Bfield}) one can re-write $C_2$ using the 
coordinates introduced in this section. The final expression 
is, however, even more complicated than (\ref{eq:Bfield}) and 
we will not give it here since (as we will argue in the next section) 
the RR form $C_2$ does not contribute to the WZ action and therefore is 
irrelevant for the present discussion. 

We will later need the expression for the RR $6$-form $C_6$. The corresponding 
$7$-form field strength is defined by:

\begin{equation}   \label{eq:F7}
F_7 =  \star_{10} F_3 - C_4 \wedge H_3.
\end{equation}
The equation of motion of $C_2$ implies that $\diff F_7=0$. 
Here $\star_{10}$ denotes the $10d$ Hodge dual. 
Denoting by $\star_6$ the Hodge dual on  $\mathnormal{M}_6$
we may re-write the first term in (\ref{eq:F7}) as:

\begin{equation}
\star_{10} F_3 = h^{-1} \star_{6} F_3 \wedge \diff x_0 \wedge \ldots \wedge \diff x_3.
\end{equation}
Remarkably, the $3$-forms $F_3$ and $H_3$ in the KS background satisfy the following relations:

\begin{equation}
 \star_6  F_3  =  g_s^{-1}  H_3
\quad
\textrm{and}
\quad
 \star_6  H_3  =  - g_s  F_3.
\end{equation}
Plugging this into the definition (\ref{eq:F7}) of $F_7$ and using
the expression (\ref{eq:C4}) for $C_4$ we arrive at the conclusion
that $F_7=0$ in the KS solution.
We thus set $C_6=0$ in the rest of this paper.

\subsection{The dual gauge theory}

The dual field theory in the KS model is a 4d $\mathcal{N}=1$ 
$SU(N+M) \times SU(N)$ gauge theory with a $SU(2) \times SU(2)$
global symmetry inherited from the conifold isometries.
Here $N$ and $M$ are the numbers of the physical and the fractional D$3$ branes
respectively. $M$ is fixed by the charge of the RR 3-form,
while $N$ is encoded in the UV behavior of the 5-form (\ref{eq:5form}).
The gauge theory is coupled to two bi-fundamental
chiral multiplets $A_{i=1,2}$ and $B_{i=1,2}$, which transform as a doublet of one of
the $SU(2)$'s each and are inert under the second $SU(2)$.
The superpotential inert under the global symmetries is:

\begin{eqnarray}    \label{eq:Wcon}
W_{\textrm{conifold}} \sim \epsilon^{ij} \epsilon^{kl} \textrm{Tr} \left( A_i B_j A_k B_l \right).
\end{eqnarray}

This theory is believed to exhibit a cascade of Seiberg dualities
reducing in the deep IR to pure $SU(M)$.
There is a simple identification of the superfields $A_{i=1,2}$ and $B_{i=1,2}$
in terms of the coordinates $z_i$'s:

\begin{eqnarray}
&&
w_1 = A_1 B_1, \qquad
w_2 = A_2 B_2, \qquad
w_3 = A_1 B_2, \qquad
w_4 = A_2 B_1, \qquad
\nonumber \\
&& \qquad
\textrm{where} \qquad
w_{1,2} = \pm z_3 + i z_4 \quad
\textrm{and} \quad
w_{3,4} = z_1 \pm i z_2.
\end{eqnarray}
Since $z_4 \sim w_1 + w_2$ the natural deformation of the superpotential
(\ref{eq:Wcon}) corresponding to the embedding $z_4 = \mu$ is \cite{Ouyang:2003df} \cite{Burrington:2004id}:

\begin{eqnarray}   
W = W_{\textrm{conifold}} + \lambda Q \left( A_1 B_1 + A_2 B_2  - \mu \right) \tilde{Q}.
\end{eqnarray}
Recall that a position of a probe D$3$ brane on the conifold is
encoded in the vacuum expectation values of the fields $A_{1,2}$ and $B_{1,2}$.
We see that the superpotential implies that the fundamentals
(arising as $3-7$ strings) become massless, when the probe D$3$ brane 
is on the D$7$ brane locus and therefore $z_4-\mu=0$ as expected. 
For the motivation of this deformation using the theory of D$3$
on $\mathcal{C}^2 / \mathcal{Z}_2$ see \cite{Ouyang:2003df}.

\section{$\kappa$-symmetry}
\label{section:Kappa}

As mentioned in the Introduction to get a supersymmetric 
brane embedding the Killing spinor $\varepsilon$ of the KS solution has to satisfy 
the $\kappa$-symmetry condition \cite{Cederwall:1996pv}, \cite{Cederwall:1996ri}, 
\cite{Bergshoeff:1996tu}, \cite{Bergshoeff:1997kr}:

\begin{equation}   \label{eq:KappaCondition}
\Gamma_{\kappa} \varepsilon = \varepsilon,
\end{equation}
where the matrix $\Gamma_{\kappa}$ depends on the geometry of the embedding as well as on
the world-volume gauge fields of the brane.
It can be expressed in the following way:

\begin{equation}
\Gamma_{\kappa} = e^{-\frac{a}{2}} \Gamma_{(0)}^\prime e^{\frac{a}{2}},
\end{equation}
where $a$ contains all the dependence on the $B$-field and the world-volume gauge field. 
More precisely it depends on the modified $2$-form field strength 
$\mathcal{F} = \varphi^\star(B) + 2 \pi \alpha^\prime F $,
where $\varphi^\star(B)$ is the pullback of the $B$-field.
The matrix $\Gamma_{(0)}^\prime$ depends on the D$p$ brane embedding 
and for $p=7$ it is given by:

\begin{equation}   \label{eq:Gamma0}
\Gamma_{(0)}^\prime = i \sigma_2 \otimes  \frac{1}{8 ! \sqrt{\vert g \vert}} \epsilon^{i_1 \ldots i_8}
      \p_{i_1} X^{n_1} \cdot \ldots \cdot \p_{i_8} X^{n_8}   \gamma_{n_1 \ldots n_8}
\qquad
\textrm{with}
\qquad      
\gamma_n=E_n^a \Gamma_a.
\end{equation}
Here $\gamma_n$ and $\Gamma_{a}$'s are the $10d$ curved and flat 
space gamma-matrices respectively and $\vert g \vert$ denotes 
the determinant of the induced metric.

The $10d$ supergravity spinor of the KS solution can be decomposed in the following way:

\begin{equation}   \label{eq:varepsilon}
\varepsilon = \zeta \otimes \chi,  
\end{equation}
where $\zeta$ is a four dimensional chiral spinor ($\Gamma^4 \zeta=\zeta$) 
and  $\chi$ is a six dimensional chiral spinor ($\Gamma^6 \chi=-\chi$).
It was shown in \cite{Kehagias:1998gn} that for a type IIB background of the form 
(\ref{eq:metric}), (\ref{eq:metric}) the spinor $\chi$ 
is given by the Killing spinor 
$\tilde{\chi}$ of 
the $6d$ internal space (the deformed conifold in the KS background) multiplied by a power
of the warp function $h(\tau)$:

\begin{equation}
\tilde{\chi} = h^{-\frac{1}{8}} \chi.
\end{equation}
This result continues to hold also in the presence of the
$3$-form $G_3 \equiv F_3 + \frac{i}{g_s} H_3$ as in the KS solution.
Moreover the supersymmetry associated with $\chi$ is unbroken provided that 
this $3$-form is of type $(2,1)$ and is imaginary self dual ($ \star_6 G_3 =i G_3$)
\cite{Grana:2001xn}, \cite{Gubser:2000vg}.

On substituting the spinor (\ref{eq:varepsilon}) into the 
the $\kappa$-symmetry condition (\ref{eq:KappaCondition}) 
and considering the case where the D$7$ brane spans the D$3$ world-volume coordinates in
(\ref{eq:metric}) and wraps a $4$-cycle of the internal space,
we see that the four dimensional part of the condition corresponding to the spinor 
$\zeta$ is trivially satisfied
and (\ref{eq:KappaCondition}) effectively reduces to the condition on 
a D$3$ brane wrapping the $4$-cycle. This problem was discussed in \cite{Marino:1999af},
where it was proven that in order to preserve the supersymmetry related to $\tilde{\chi}$
the solution has to satisfy the following three restrictions:

\begin{enumerate}
\item The embedding is holomorphic \cite{Becker:1995kb}.
\item The modified $2$-form field strength $\mathcal{F} = \varphi^\star(B) + 2 \pi l_s^2 F$
        is of type $(1,1)$. Here $\varphi^\star(B)$ is the pullback of the $2$-form $B$.
\item The pullback of the K\"ahler $2$-form $J$ and the form $\mathcal{F}$ satisfy the equation:
\begin{equation}   \label{eq:Strominger}
\varphi^\star (J) \wedge \mathcal{F} 
     = \tan \theta \left( \textrm{vol}_4 - \half \mathcal{F} \wedge \mathcal{F} \right), 
\end{equation}         
where $\textrm{vol}_4 = \half \varphi^\star (J) \wedge \varphi^\star (J) $ is the canonical
volume element of the $4$-cycle and $\theta$ is a constant parameter.
\end{enumerate}
In the rest of this section we will examine these conditions for the deformed conifold case.
To this end we will need the expression for the $B$-field in terms of the complex 
coordinates $z_i$'s \cite{Herzog:2001xk}:

\begin{equation}  \label{eq:so(4)}
B = b(\tau) \epsilon_{i j k l} z_i \bar{z}_j \diff z_k \wedge \diff \bar{z}_l
\qquad
\textrm{with}
\qquad
b(\tau) = \frac{i g_s M \alpha^\prime}{2 \epsilon^4} \frac{\tau \coth \tau - 1}{\sinh^2  \tau}.
\end{equation}
This expression is explicitly $SO(4)$ invariant. Moreover, in this gauge the 
$B$-field is of type $(1,1)$ and so for vanishing $F_{\mu\nu}$ and 
for an arbitrary holomorphic embedding 
the $2$-form $\mathcal{F}$ also of type $(1,1)$
in agreement with the supersymmetry restriction described above.

Now let us consider the condition (\ref{eq:Strominger}).
We want to show that the embedding $z_4 = \mu$ is a solution  
of (\ref{eq:Strominger}) for $\theta =0$. We will assume also that $F_{\mu\nu}=0$.
Plugging $z_4 = \mu$ into (\ref{eq:so(4)}) and expressing $z_3$ in terms of $z_1$ and $z_2$
we get:

\begin{eqnarray}
\varphi^\star (B) = \mu^2 b(\tau) & \cdot & \Big[  
           - ( z_2 - \bar{z}_2) \left( \frac{z_1}{z_3} - \frac{\bar{z}_1}{\bar{z}_3} \right) 
                          \diff z_1 \wedge \diff \bar{z}_1 
           + ( z_1 - \bar{z}_1) \left( \frac{z_2}{z_3} - \frac{\bar{z}_2}{\bar{z}_3} \right) 
                          \diff z_2 \wedge \diff \bar{z}_2
\nonumber \\                          
 &&          + \left( z_3 - \bar{z}_3 + (z_1-\bar{z}_1)  \frac{z_1}{z_3} 
                                    + (z_2-\bar{z}_2)  \frac{\bar{z}_2}{\bar{z}_3}  \right)            
                                    \diff z_1 \wedge \diff \bar{z}_2 
\nonumber \\                                   
 &&          - \left( z_3 - \bar{z}_3 + (z_1-\bar{z}_1)  \frac{\bar{z}_1}{\bar{z}_3} 
                                    + (z_2-\bar{z}_2)  \frac{z_2}{z_3}              \right)             
                                    \diff z_2 \wedge \diff \bar{z}_1  
                    \Big]                                                                         
\end{eqnarray}
Similarly the pullback of the K\"ahler $2$-form is:

\begin{eqnarray}
\varphi^\star (J) &=& \left. \p_\alpha \bar{\p}_\beta \mathcal{F} \diff z_\alpha \wedge \diff \bar{z}_\beta 
                                   \right\vert_{z_4 = \mu} =
\nonumber \\
       &=&    \mathcal{F}^{\prime\prime}   \left(  
                            2 \left\vert z_1 \right\vert^2 \diff z_1 \wedge \diff \bar{z}_1 
                         +  2 \left\vert z_2 \right\vert^2 \diff z_2 \wedge \diff \bar{z}_2 
                         +   z_1 \bar{z}_2 \diff z_1 \wedge \diff \bar{z}_2  
                         +   z_2 \bar{z}_1 \diff z_2 \wedge \diff \bar{z}_1    
                                   \right)
\nonumber \\
           && + \mathcal{F}^\prime \Bigg(  
                          \left( 1 + \frac{\vert z_1 \vert^2}{\vert z_3 \vert^2} \right)  
                                                       \diff z_1 \wedge \diff \bar{z}_1 
                       +  \left( 1 + \frac{\vert z_2 \vert^2}{\vert z_3 \vert^2} \right)  
                                                       \diff z_2 \wedge \diff \bar{z}_2
\nonumber \\                                                       
           && \qquad \qquad
                      +   \frac{z_1 \bar{z}_2}{\vert z_3 \vert^2}  \diff z_1 \wedge \diff \bar{z}_2  
                      +   \frac{z_2 \bar{z}_1}{\vert z_3 \vert^2}  \diff z_2 \wedge \diff \bar{z}_1    
                                   \Bigg) .                                        
\end{eqnarray}
This is now a straightforward exercise to verify that $\varphi^\star(J) \wedge \varphi^\star(B)=0$
and thus the condition (\ref{eq:Strominger}) is fulfilled
for the embedding $z_4=\mu$. There is, however,
a simple symmetry argument one can use to significantly simplify the proof.
Both the $B$-field and the K\"ahler form are $SO(4)$ invariant objects. In particular,
they are invariant under rotations in the $(z_1,z_2)$ plane. On the other hand,
the K\"ahler form is also inert under

\begin{equation}   \label{eq:reflection}
z_1 \to z_2  \qquad z_2 \to z_1,
\end{equation}
while the $B$-field transforms under this reflection as $B \to - B$.
Furthermore, we have $\varphi^\star(J) \wedge \varphi^\star(B)=
\phi \cdot \diff z_1 \wedge \diff z_2 \wedge\diff \bar{z}_1 \wedge\diff {z}_2$
for some function $\phi = \phi ( z_1,z_2,\bar{z}_1,\bar{z}_2 )$.
Since the embedding $z_4 = \mu$ is both invariant under rotations and the reflection
(\ref{eq:reflection}) we conclude that $\phi$ is invariant under rotations 
and transforms as $\phi \to -\phi$ under (\ref{eq:reflection}). The only functions 
with these two properties is $\phi \equiv 0$.

Remarkably, this proof is valid for a general holomorphic embedding of the form:

\begin{equation}
z_4 = f \left(z_1^2+z_2^2 \right).
\end{equation}
In particular, the configurations (\ref{eq:Karch}) and (\ref{eq:Ouyang})
are of this form and therefore satisfy the $\kappa$-symmetry condition as was
announced in the Introduction.

\section{Spectrum of mesons}
\label{section:DBIWZ}

In this section we will study the quadratic fluctuations around the
D$7$ brane configuration $z_4 = \mu$ for $\mu=0$.
The set of dynamical fields on the brane includes $8d$ gauge potentials
and two $8d$ scalars corresponding to the fluctuations along the two
transverse directions. 
We will consider the abelian case $N_{\textrm{f}}=1$ and will examine 
only the lowest KK modes, namely we will assume that the $8d$ fields
depend only on the radial coordinate $\tau$ and on the D$3$ world-volume coordinates $x_\mu$'s.
Fields with a non-trivial dependence on the angular coordinates associated with the $1$-forms
$h_i$'s carry non-zero spins and charges related to the $SU(2)$ isometry of
the embedding and therefore have no counterparts in the dual QCD.

As we mentioned in the Introduction the masses of the fluctuation modes
give the spectrum of the gauge theory mesons. There are two kinds of mesons.
The gauge fields $A_\mu$ with non-vanishing components along the non-compact 
directions $x_\mu$ correspond to vector mesons in the gauge theory, 
while the gauge fields directed along the compact angular directions $(A_1,A_2,A_3)$
and the scalar fields on the brane correspond to scalar mesons.

The D$7$ brane action consists of two parts:

\begin{equation}
S = S_{\textrm{DBI}} + S_{\textrm{WZ}},
\end{equation}
where 

\begin{equation}
S_{\textrm{DBI}} = - \mu_7 e^{-\phi_0} \int \diff \sigma^8 
       \sqrt{\varphi^\star(g) + \mathcal{F}},
\quad
\textrm{and}
\quad
S_{\textrm{WZ}} = \mu_7 \int \sum_p C_{p+1} \wedge e^{\mathcal{F}}.
\end{equation}
Here $\mu_7 = (2 \pi)^{-7} l_s^8$ , $e^{-\phi_0} = g_s^{-1}$, $\varphi^\star(g)$
is the pullback of the $10d$ metric (\ref{eq:metric}) and 
$\mathcal{F} = \varphi^\star(B) + 2 \pi l_s^2 F$ is the modified
$2$-form field strength as in the previous section. We saw in Section \ref{section:KS}
that $C_6=0$ in the KS model. Moreover, $C_0$ also vanishes in this background and
$C_2$ has no legs along the $4d$ space-time spanned by the coordinates $x_\mu$'s.
Thus we conclude that the only contribution to the WZ part of the action is due to the 
RR $4$-form $C_4$ (\ref{eq:C4}).

It appears that up to the quadratic terms the D$7$ brane action takes the following form:

\begin{eqnarray}     \label{eq:theaction}
S &=& - \mu_7 \int \diff^4 x \diff \tau h_1 h_2 h_3 \times
\nonumber \\
&&  \times   \Bigg[ \sqrt{ -\vert g_{(0)} \vert} \Bigg(
                   1 + \frac{1}{4} \left( 1+ \frac{1}{\cosh \tau} \right) h^{\half} A^2 
                                                g_{(0)}^{\mu \nu} \p_\mu \gamma \p_\nu \gamma 
                      + \frac{1}{4} \left( \frac{A}{B} \right)^2  \left(\p_\tau \gamma \right)^2  -                          
  \nonumber \\                                                
  &&          \quad        - \frac{1}{4}  \left( 1 - \frac{1}{4} \left( \frac{A}{B} \right)^2 
                                       \left( 1 - \frac{1}{\cosh \tau} \right) \right) \gamma^2 
                     +\frac{1}{4} g_{(0)}^{IJ} g_{(0)}^{KL} \mathcal{F}_{IK} \mathcal{F}_{JL} 
                                       \Bigg)  -
   \nonumber \\                                                
  &&         \qquad    - \frac{1}{2!} h^{-1} \epsilon^{ijkl} \mathcal{F}_{ij} \mathcal{F}_{kl}
      \Bigg] .                                                                 
\end{eqnarray}
Here $\mu$, $\nu, \ldots$ refer to the $4d$ non-compact world-volume coordinates, 
$I,J,\ldots$ are the $8d$ indices, 
$i,j,\ldots = \tau,1,2,3$ denote the radial and the internal space indices,  
the functions $A(\tau)$ and $B(\tau)$ are given in (\ref{eq:AB}) and the $8d$ metric $g_{(0)}^{IJ}$ is:

\begin{equation}      \label{eq:8dmetric}
g_{(0)}^{IJ} = \textrm{diag} \left( -h^\half,h^\half,h^\half,h^\half, 
           \frac{ \epsilon^{-\frac{4}{3} } h^{-\half}}{B^2}, 
           \frac{ \epsilon^{-\frac{4}{3} } h^{-\half}}{2 A^2 \left( 1- \frac{1}{\cosh \tau} \right)},  
           \frac{ \epsilon^{-\frac{4}{3} } h^{-\half}}{2 A^2 \left( 1+ \frac{1}{\cosh \tau} \right)}, 
           \frac{ \epsilon^{-\frac{4}{3} } h^{-\half}}{4 B^2}
           \right).
\end{equation}  
It implies that $\sqrt{ -\vert g_{(0)} \vert} = 4 A^2(\tau) B^2(\tau) \tanh \tau$.
The components of the $2$-form $\mathcal{F}$ appearing in the action (\ref{eq:theaction}) are:

\begin{eqnarray}
&&      
        \mathcal{F}_{\tau 1} = 2 \pi l_s^2 F_{\tau 1}   \qquad
        \mathcal{F}_{23} = 2 \pi l_s^2 F_{23}           \qquad
        \mathcal{F}_{\tau 3} = 2 \pi l_s^2 F_{\tau 3}   \qquad
        \mathcal{F}_{12} = 2 \pi l_s^2 F_{12}
\nonumber \\
&&      
        \mathcal{F}_{\tau 2} = - \frac{1}{4} M l_s^2 (k(\tau)-f(\tau)) (\cosh \tau+1) \gamma^\prime 
                               + 2 \pi l_s^2 F_{\tau 2}         
\nonumber \\
&&
        \mathcal{F}_{13} = \frac{1}{4} M l_s^2 (k(\tau)-f(\tau)) (\cosh \tau-1) \gamma
                               + 2 \pi l_s^2 F_{13}   
\nonumber \\
&&
        \mathcal{F}_{\mu 1} = 2 \pi l_s^2 F_{\mu 1}   \qquad
        \mathcal{F}_{\mu 3} = 2 \pi l_s^2 F_{\mu 3}  
\nonumber \\
&&
        \mathcal{F}_{\mu 2} = \frac{1}{4} M l_s^2 (k(\tau)-f(\tau)) (\cosh \tau+1) \p_\mu \gamma
                               + 2 \pi l_s^2 F_{\mu 2}                                              
\end{eqnarray}
In writing these components we used (\ref{eq:Bfield}).
We see that the action (\ref{eq:theaction})
has no terms linear in the fields. Hence the supersymmetric configuration $z_4=0$
solves the probe equations of motion. It is interesting to notice that the field
$\delta(x_\mu,\tau)$ does not appear in the action. A simple check reveals that 
the term proportional to $\delta^2$ will be produced in the $\mu \neq 0$ case.

From the expressions for $\mathcal{F}_{13}$
and $\mathcal{F}_{\tau 2}$ it is evident that 
there is a mixing between the scalar field $\gamma$ and the $A_2$ component of the gauge field.

In the rest of this section we will use the action (\ref{eq:theaction}) to calculate
the masses of $4d$ vector and scalar mesons.
In what follows we will impose the gauge $A_\tau=0$.

\subsection{Vector mesons}

For the vector field polarized along the non-compact world-volume coordinates
we will adopt the ansatz $A_\mu(x, \tau) = v_\mu e^{i k \cdot x} a(\tau)$,
where $-k^2 = M_{\textrm{m}}^2$ denotes the $4d$ mass.
The equation of motion extracted from (\ref{eq:theaction}) 
is:

\begin{equation}
\p_c \left( \sqrt{g} g^{cd} g^{ab} F_{db} \right) = 0,
\end{equation}
where the indices $a,b,\ldots$ run over the radial $\tau$
and the $4d$ coordinates $x_\mu$ and $g_{ab}$ corresponds to an appropriate metric component
in (\ref{eq:8dmetric}).
It implies
that $v \cdot k =0$ and the function $a(\tau)$ satisfies:

\begin{equation}
\left( A^2(\tau) \tanh (\tau) a^\prime (\tau) \right)^\prime 
   +\lambda_n^2  I(\tau) A^2(\tau) B^2(\tau) \tanh (\tau) a (\tau) = 0.
\end{equation}
Here $\lambda_n = M_{\textrm{m}} / m_{\textrm{gb}}$ with $m_{\textrm{gb}}$ being the scale mass of 
the gauge theory glueballs (\ref{eq:gb}) and $I(\tau)$ is defined in (\ref{eq:I}). 
The normalizable solution of the above differential equation at $\tau \to \infty$
is $a(\tau) \sim e^{-2 \tau/3}$. On the other hand, the converging solution 
near $\tau=0$ is $a(\tau) \sim \textrm{const}$.
Matching numerically the regular solutions in the UV ($\tau \to \infty$) and in IR ($\tau \to 0$) 
results in a set of discrete values
of the parameter $\lambda_n$.
This computation leads to the following result 
for the lowest masses of vector mesons in the dual gauge theory:

\begin{equation}    \label{eq:amu}
\lambda_n = \frac{M_{\textrm{m}}}{m_{\textrm{gb}}} = 4.32, \, 5.81, \, 7.32, \, 8.85, \, 10.39,  \ldots. 
\end{equation}

\subsection{Scalar mesons}

We next consider the gauge field $A_1$ and $A_3$.
Substituting $A_1(x, \tau) = a_1(\tau) e^{i k \cdot x}$ and 
$A_3(x, \tau) = a_3(\tau) e^{i k \cdot x}$ into the action 
we end up the following equations for $a_{1,3}(\tau)$:

\begin{equation}
\left (  \frac{\coth \left( \frac{\tau}{2} \right)}{I(\tau)} a_1^\prime (\tau) \right)^\prime =
     \left[ \frac{1}{2}\left( \frac{1}{I(\tau)} \right)^\prime + 
            \frac{1}{4} \frac{\tanh \left( \frac{\tau}{2} \right)}{I(\tau)}  
            - \lambda_n^2 B^2(\tau) \coth \left( \frac{\tau}{2} \right) \right] a_1(\tau)
\end{equation}
and
\begin{equation}
\left( \frac{A^2(\tau)}{I(\tau) B^2(\tau)} a_3^\prime (\tau) \right)^\prime =
     \left[ \left( \frac{1}{I(\tau) } \right)^\prime + 
             \frac{B^2(\tau)}{I(\tau) A^2(\tau)} \coth (\tau) 
              - \lambda_n^2 A^2(\tau) \tanh (\tau) \right] a_3(\tau).
\end{equation}
Now the regular solutions at $\tau \to \infty$ are $a_1(\tau) \sim e^{-11/6 \tau}$
 and  $a_3(\tau) \sim e^{-2 \tau}$, while at $\tau \to 0$ the converging solutions are 
$a_1(\tau) \sim \tau^2$ and $a_3(\tau) \sim \tau$.
This time the "shooting" technique yields the following spectra:

\begin{equation}  \label{eq:a1}
\lambda_n=\frac{M_{\textrm{m}}}{m_{\textrm{gb}}} = 3.38, \, 4.92, \, 6.38, \, 7.87, \, 9.37, \ldots 
\end{equation}
and
\begin{equation}  \label{eq:a3}
\lambda_n=\frac{M_{\textrm{m}}}{m_{\textrm{gb}}} = 4.18, \, 5.67, \, 7.17, \, 8.71, \, 10.28, \ldots 
\end{equation}
for the mesons related to the field $A_1$ and $A_3$ respectively.

Finally let us consider the fields 
$\gamma$ and $A_2$. Following the same steps as in the previous discussion
we we will assume that $\gamma=\gamma(\tau) e^{i k \cdot x}$
and $A_2=a_2(\tau) e^{i k \cdot x}$.
We arrive to a set of two 2nd order differential equations for $\gamma(\tau)$
and $a_2(\tau)$:

\begin{eqnarray}
&&
 \left( I^{-1}  \tanh \frac{\tau}{2} 
         \left( -(k -f) (\cosh \tau +1 ) \gamma^\prime +\tilde{a}_2^\prime \right) +
        \frac{1}{2} I^{-1}  \left( (k -f) (\cosh \tau -1 ) \gamma - \tilde{a}_2 \right)
 \right)^\prime = 
\nonumber \\
&& \qquad 
 =\frac{1}{4} I^{-1} \coth \frac{\tau}{2} \left( -(k -f) (\cosh \tau -1 ) \gamma +\tilde{a}_2 \right) +
 \frac{1}{2} I^{-1} \left( (k -f) (\cosh \tau -1 ) \gamma^\prime  - \tilde{a}_2^\prime \right)- 
\nonumber \\
 && \qquad \qquad \qquad
 - \lambda_n^2 \left( \frac{B}{A} \right)^2 \tanh \frac{\tau}{2}  
                       \left( (k -f) (\cosh \tau +1 ) \gamma +\tilde{a}_2 \right)                                                                 
\end{eqnarray}
and 

\begin{eqnarray}
&&
 \bigg( I^{-1}  (k-f)  \Big( (\cosh \tau +1) \tanh \frac{\tau}{2} 
         \left( (k -f) (\cosh \tau +1 ) \gamma^\prime - \tilde{a}_2^\prime \right) +
\nonumber \\
&& \qquad         
       + \frac{1}{2}  (\cosh \tau -1) \left( -(k -f) (\cosh \tau -1 ) \gamma + \tilde{a}_2 \right) \Big) +
        2^{\frac{14}{3}} A^4 \tanh \tau \, \gamma^\prime
 \bigg)^\prime = 
\nonumber \\
&&  
 = \frac{1}{2} I^{-1} (k-f) (\cosh\tau -1) \bigg( \frac{1}{2} \coth \frac{\tau}{2} 
            \big( (k -f) (\cosh \tau -1 ) \gamma - \tilde{a}_2 \big) -
\nonumber \\
&& \qquad
 - \left( (k -f) (\cosh \tau -1 ) \gamma^\prime - \tilde{a}_2^\prime \right) \bigg) -                    
\nonumber \\
 && \,
 - \lambda_n^2 B^2 (k-f) (\cosh\tau +1) \tanh \frac{\tau}{2}  \big( (k -f) (\cosh \tau +1 ) \gamma +\tilde{a}_2 \big) -                                                               
\nonumber \\
 && \,
 - 2^{\frac{14}{3}} A^2 B^2 \tanh \tau \left( -1 + \frac{1}{4} \left( \frac{A}{B} \right)^2
                                                  \left( 1 - \frac{1}{\cosh \tau} \right) -
                                    \lambda_n^2 I A^2 \left( 1 - \frac{1}{\cosh \tau} \right)  \right) \gamma.
\end{eqnarray}
Here $\tilde{a}_2 \equiv \frac{8 \pi}{M} a_2$.
The regular solution at $\tau \to \infty$ behaves as 
$\left( \tau \gamma(\tau) - \tilde{a}_2 \right) \sim e^{-11/6 \tau}$
and at $\tau \to 0$ we have 
$\gamma(\tau), a_2(\tau) \sim \textrm{const}$.
Solving numerically this set we get:

\begin{equation}
\lambda_n=\frac{M_{\textrm{m}}}{m_{\textrm{gb}}} = 4.01, \, 5.52, \ldots.
\end{equation}

\section{Discussion}
\label{section:discussion}

In this paper we have investigated holomorphically embedded 
D$7$ branes in the KS background in order to get a supergravity description
of $\mathcal{N}=1$ QCD with flavors. We considered the $N_{\textrm{f}}=1$ case
and ignored the back-reaction of the D$7$ brane probe on the supergravity background.
Studying the $\kappa$-symmetry condition derived in \cite{Marino:1999af}
we argued that our brane configuration $z_4=\mu$ preserves the background supersymmetry.
We also gave a simple criterion to verify whether holomorphic embeddings suggested by other authors
are supersymmetric. The embedding $z_4=\mu$ discussed in this paper preserves the $SU(2)$ 
subgroup of the deformed conifold isometries. As we have shown this fact significantly 
simplifies the choice of the coordinates along the D$7$ brane and transversal to it.
Using this coordinate parameterization we calculated the spectrum of the vector and scalar mesons
for the special $\mu=0$ case.
To this end we studied the equations of motion
extracted from the quadratic fluctuations around the probe.
The meson masses were computed this way by numerical matching 
of the regular solutions in the UV and in the IR.
We found that the spectrum satisfies the following important properties:

\begin{itemize}
\item
It exhibits a mass gap
\item
The masses are of the order of the glueball mass $m_{\textrm{gb}}$.
\item
The vector mesons are heavier than the corresponding scalar mesons.
Remarkably, this property is not shared by the meson spectrum found
in \cite{Sakai:2003wu}.
\item
At least for the lowest levels $n=1, \ldots,5$ we have obtained here,
the masses $M_{\textrm{m}}$ of the vector and the scalar mesons grow linearly as a function of $n$.
In all the cases (\ref{eq:amu}), (\ref{eq:a1}) and (\ref{eq:a3}) we have found that:

\begin{equation}
M_{\textrm{m}} (n) = \Delta m \cdot n + \textrm{const},
\end{equation}
where $\Delta m \approx 1.5 m_{\textrm{gb}}$.
\item
For $\mu=0$ there is no equation associated with the scalar field $\delta$,
since this field does not contribute to the quadratic fluctuations of the probe.
This is very similar to the results of \cite{Sakai:2003wu}. This feature certainly has
to do with the fact that for $\mu=0$ there is a residual $U(1)$ isometry along the D$7$ probe brane
as we have mentioned in Section \ref{section:KS}.
\end{itemize}

There are plenty of open problems to be explored:
 
In this paper
we computed the meson mass spectrum only for the $\mu=0$ case.
It will be very interesting to find the spectrum for $\mu>0$.
This spectrum will be characterized by two mass scales: $\mu$ and the glueball mass
related to the conifold parameter $\epsilon$. 
Although we saw that the embedding $z_4=\mu$ is supersymmetric
for any $\mu$, it might still be possible that this configuration
does not solve the probe equations of motion.
It is also important to perform the spectrum computation for
other holomorphic configurations like (\ref{eq:Karch})
and (\ref{eq:Ouyang}) and to compare it to the results of this paper.

In \cite{Kuperstein:2003yt} an analytic first order deformation of the KS background
was found. From the dual gauge theory point of view this deformation describes 
supersymmetry soft breaking gaugino mass terms.
It will be interesting to consider a D$7$ brane probing the non-supersymmetric background
and to understand how the deformation
modifies the meson spectrum we have obtained here.

Finally, it is very desirable to find a modification of the KS
solution which incorporates fully localized D$7$ branes.
This is needed for studying gauge theories where 
the approximation $N_{\textrm{f}} \ll N_{\textrm{c}}$ is not
justified. First steps in this direction have been made in
\cite{Ouyang:2003df} and \cite{Burrington:2004id}.

\acknowledgments

The author would like to thank
Yaron Oz, Ofer Aharony and Cobi Sonnenschein for fruitful discussions.
This work was supported in part by the German-Israeli Foundation for
Scientific Research and by the Israel Science Foundation.

\appendix

\section{The explicit expressions for the $1$-forms $h_i$ and $\tilde{h}_i$}
\label{01}

Here we give the expressions for the form $h_i$, $\tilde{h}_i$
($i=1,2,3$) in terms of the five angular coordinates $\theta_1$, $\phi_1 $,
$\theta_2$, $\phi_2 $ and $\psi$:

\begin{eqnarray}
h_1 &=& - \cos \frac{\psi}{2} \sin \theta_1 \diff \phi_1 + \sin \frac{\psi}{2} \diff \theta_1
\nonumber  \\
h_2 &=& - \sin \frac{\psi}{2} \sin \theta_1 \diff \phi_1 - \cos \frac{\psi}{2} \diff \theta_1
\nonumber  \\
\tilde{h}_1 &=& - \cos \frac{\psi}{2} \sin \theta_2 \diff \phi_2 + \sin \frac{\psi}{2} \diff \theta_2
\nonumber  \\
\tilde{h}_2 &=& - \sin \frac{\psi}{2} \sin \theta_2 \diff \phi_2 - \cos \frac{\psi}{2} \diff \theta_2
\nonumber  \\
h_3 + \tilde{h}_3 &=&  \diff \psi  + \cos \theta_1 \diff \phi_1 + \cos \theta_2 \diff \phi_2.
\end{eqnarray}

\section{The connection between the forms $g_{i=1, \ldots,5}$ and the 
         forms $h_{i=1,2,3}$, $\tilde{h}_{i=1,2,3}$}
\label{02}

\begin{eqnarray}
\left(
\begin{array}{c}
h_1 \\ h_2
\end{array}
\right)
&=&
\left(
\begin{array}{cc}
   \cos \frac{\psi}{2}  & \sin \frac{\psi}{2} \\
   \sin \frac{\psi}{2}  & -\cos \frac{\psi}{2}  
\end{array}
\right)
\left(
\begin{array}{c}
  \frac{1}{\sqrt{2}} (g_1 + g_3) \\ 
  \frac{1}{\sqrt{2}} (g_2 + g_4) 
\end{array}
\right)
 \nonumber \\
\left(
\begin{array}{c}
\tilde{h}_1 \\ \tilde{h}_2
\end{array}
\right)
&=&
\left(
\begin{array}{cc}
   -\cos \frac{\psi}{2}  & -\sin  \frac{\psi}{2} \\
   \sin \frac{\psi}{2} &  -\cos   \frac{\psi}{2}  
\end{array}
\right)
\left(
\begin{array}{c}
  \frac{1}{\sqrt{2}} (g_3 - g_1) \\ 
  \frac{1}{\sqrt{2}} (g_4 - g_2) 
\end{array}
\right)
\nonumber \\
\textrm{and}  \qquad &&
h_3 + \tilde{h}_3 = g_5.
\end{eqnarray}

\bibliography{D7}

\end{document}

%% file: D7KS1.pstex_t
\begin{picture}(0,0)%
\includegraphics{D7KS1.pstex}%
\end{picture}%
\setlength{\unitlength}{2368sp}%
\begingroup\makeatletter\ifx\SetFigFont\undefined%
\gdef\SetFigFont#1#2#3#4#5{%
  \reset@font\fontsize{#1}{#2pt}%
  \fontfamily{#3}\fontseries{#4}\fontshape{#5}%
  \selectfont}%
\fi\endgroup%
\begin{picture}(2424,4365)(1189,-4414)
\put(2401,-4336){\makebox(0,0)[b]{\smash{{\SetFigFont{10}{12.0}{\rmdefault}{\mddefault}{\updefault}$\tau=0$}}}}
\put(2889,-1474){\makebox(0,0)[rb]{\smash{{\SetFigFont{10}{12.0}{\rmdefault}{\mddefault}{\updefault}D$7$}}}}
\end{picture}%

%% file: D7KS2.pstex_t
\begin{picture}(0,0)%
\includegraphics{D7KS2.pstex}%
\end{picture}%
\setlength{\unitlength}{2368sp}%
\begingroup\makeatletter\ifx\SetFigFont\undefined%
\gdef\SetFigFont#1#2#3#4#5{%
  \reset@font\fontsize{#1}{#2pt}%
  \fontfamily{#3}\fontseries{#4}\fontshape{#5}%
  \selectfont}%
\fi\endgroup%
\begin{picture}(2497,4365)(7651,-4414)
\put(7651,-2911){\makebox(0,0)[rb]{\smash{{\SetFigFont{10}{12.0}{\rmdefault}{\mddefault}{\updefault}$\tau=\tau_{\textrm{min}}$}}}}
\put(8926,-4336){\makebox(0,0)[b]{\smash{{\SetFigFont{10}{12.0}{\rmdefault}{\mddefault}{\updefault}$\tau=0$}}}}
\put(9451,-1561){\makebox(0,0)[rb]{\smash{{\SetFigFont{10}{12.0}{\rmdefault}{\mddefault}{\updefault}D$7$}}}}
\end{picture}%